\documentclass[preprints,article,accept,moreauthors]{Definitions/mdpi} 
\firstpage{1} 
\pubvolume{xx}
\issuenum{1}
\articlenumber{1}
\pubyear{2020}
\copyrightyear{2020}
\history{This version: \today; Received: date; Accepted: date; Published: date}

\usepackage{epsfig}
\usepackage[percent]{overpic}
\usepackage{ulem}

\newcommand{\f}[2]{\frac{#1}{#2}}

\renewcommand{\vec}[1]{\mathbf{#1}}

\usepackage{psfrag}

\usepackage{xspace}
\newcommand{\Rc}{\ensuremath{\mathrm{\hat{R}_c}}}
\newcommand{\R}{\ensuremath{\hat{\mathrm{R}}}}

\allowdisplaybreaks
\Title{Onset of Inertial Magnetoconvection in Rotating Fluid Spheres}

\Author{Radostin D. Simitev$^{1,}$*\orcidA{}, Friedrich H. Busse$^{2}$\orcidB{}}

\AuthorNames{Radostin Simitev and Friedrich Busse}

\address{%
$^{1}$ \quad School of Mathematics and Statistics, University of Glasgow -- Glasgow G12 8QQ, UK\\
$^{2}$ \quad Institute of Physics,  University of Bayreuth -- 95440 Bayreuth,  Germany}

\corres{Correspondence: Radostin.Simitev@glasgow.ac.uk}

\abstract{
The onset of convection in the form of magneto-inertial  waves in a
rotating  fluid sphere permeated by a constant axial electric current
is studied through a perturbation analysis.   Explicit expressions for
the dependence of the Rayleigh number on the azimuthal wavenumber are
derived in the limit of high thermal diffusivity.  Results for the
case of thermally infinitely conducting boundaries and for the case of
nearly thermally insulating boundaries are obtained.
}

\begin{document}

%
%
%
%
%
%
%

\section{Introduction}

Buoyancy-driven motions of rotating, electrically conducting fluids
in the presence of magnetic fields represent a fundamental aspect of
the dynamics of stellar and planetary interiors,  
e.g.~\citep{Jones2011,Roberts2013,Charbonneau2014,Ogilvie2016}.
The problem of magnetic field generated and sustained by convection is
rather difficult to attack both analytically and numerically because of its
essential nonlinearity and scale separation
\citep{Glatzmaier2002,Miesch2015}. Valuable insights can be 
gained by studying magnetoconvection, the simpler case of an imposed
magnetic field, which has received much attention ever since the early
work of \citet{chandrasekhar1961hydrodynamic}, see \citep{ZhangSchubert2000,Weiss2014}.  
For instance, propagation of rotating magnetoconvection modes excited
in the deep convective region of the Earth's core has been proposed as 
a possible mechanism for explaining features of observed longitudinal
geomagnetic drifts \citep{Hide1966,Malkus1967}, see also the recent
review of \citet{Finlay2010}. A rather detailed classification of
magnetoconvection waves in a rotating cylindrical annulus has been recently
attempted by \citet{Hori2014} and \cite{Hori2015} who
proceeded further to make useful comparisons with nonlinear spherical
dynamo simulations and to provide estimates for the strength of the
“hidden” azimuthal part of the magnetic field within the core.  These
authors used the rotating annulus model of Busse \cite{Busse1970,Busse1986}
and only considered values of the Prandtl 
number of the order unity. However, both spherical geometry as
well as small values of the Prandtl number are essential features of a
planetary or a stellar interior \citep{Simitev2005a}.  
At sufficiently small values of the Prandtl number a different style
of convection exists that is sometimes called inertial or
equatorially-attached convection or thermo-inertial waves  
\citep{ZhangBusse1987,Ardes1997,Simitev2003,Plaut2005}. In this limit
convection oscillates so fast that the viscous force does not enter
the leading-order balance. The latter is then reduced to the Poincar\'e
equation in a rotating spherical system
\citep{Zhang1994,Zhang1995,ZhangLiao2017}. On the longer time scale
of the next order of approximation the buoyancy force maintains
convection against the weak viscous dissipation. This regime of
convection thus represents a transition between thermal convection and
wave propagation in rapidly rotating geometries.  It is important to
understand how this regime of inertial convection is affected by an
imposed magnetic field. 

\looseness=-1
With this in mind, we study in the present paper the onset of
magneto-inertial-convection. In particular, we consider a rotating fluid sphere
permeated by a constant axial electric current as proposed by
\citet{Malkus1967} in the limit of low viscosity and high thermal
diffusivity (small Prandtl number). 
A similarly configured problem was also investigated by
\citet{ZhangBusse1995} who derived an explicit dependence of the
critical Rayleigh number on the imposed field strength but were not
able to obtain an explicit dependence on the azimuthal wavenumber of
the modes since this requires the evaluation of a volume integral of
the temperature perturbation. For this reason \citet{ZhangBusse1995} were unable to investigate
the competition of modes and determine the actual critical parameters
for the onset of convective motion. In an earlier paper
we proposed a Green's function method for the exact solution of the
heat equation \citep{BusseSimitev2004} which then allowed the
analytical evaluation of the integral quantities needed to find a
fully explicit expressions for the critical Rayleigh number and
frequency for the onset of convection and to study mode competition. 
Here we apply the same approach to the case of magneto-inertial
convection and we consider both value and flux boundary conditions for
the temperature. 

In the following we start with the mathematical formulation of the
problem in section 2. The special limit of a high ratio of thermal to
magnetic diffusivity will be treated in section 3. The general case
requires the symbolic evaluation of lengthy analytical expressions and
will be presented in section 4. A discussion of the results and an
outlook on related problems will be given in the final 
section 5 of the paper. 

\begin{figure}[t]
\begin{center}
\hspace*{10mm}
\includegraphics[height=0.4\textwidth,clip=true]{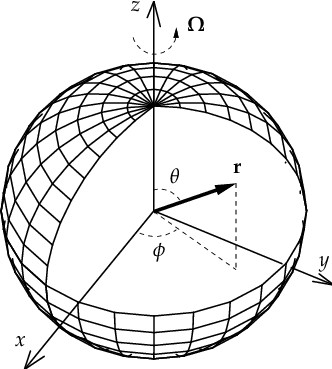}
\end{center}
\caption[]{Geometrical configuration of the problem. A part of the
 outer spherical surface is removed to expose the interior of sphere
 to which the conducting fluid is confined.} 
\label{f.01}
\end{figure}

\section{Mathematical formulation of the problem}

We consider a homogeneously heated and self-gravitating sphere as illustrated
in figure \ref{f.01}. The sphere is filled with incompressible and electrically
conducting fluid characterized by its magnetic diffusivity $\eta$,
kinematic viscosity $\nu$, thermal diffusivity $\kappa$ and density
$\varrho$. 
The sphere is rotating with a constant angular velocity $\Omega\vec k$ 
where $\vec k$ is the axial unit vector. The gravity field is given by
$\vec g = - g r_0\vec r$ where $\vec r$ is the position vector with
respect to the centre of the sphere, $r$ is its length measured in
fractions of the radius $r_0$ of the sphere and $g$ is the amplitude
of the gravitational acceleration.  
Following \citet{Malkus1967}, we assume that the fluid sphere is
permeated by a toroidal magnetic field $\vec B \sim \vec
k\times \vec r$.   
Since the Lorentz force like the centrifugal force can be
balanced by the pressure gradient a static state of no motion exists
with the temperature distribution $T_S = T_0 - \beta r_0^2 r^2 /2$.
We employ the Boussinesq approximation and assume constant material
properties $\eta$, $\nu$, $\kappa$, and $\varrho$ everywhere except in
the buoyancy term where the density is assumed to have a linear
dependence on temperature with a coefficient of thermal expansion
$\alpha \equiv (d \varrho/dT)/\varrho =$ const. 

In order to study the onset of magnetoconvection in this system we
consider the linearized momentum, magnetic induction, heat,
continuity, and solenoidality equations, 
\begin{subequations}
\label{100}
\begin{gather}
\label{102}
\partial_t \tilde{\vec{u}} + \tau \vec k \times
\tilde{\vec u} + \nabla (\pi-\tilde{\vec b}\cdot \vec j\times \vec r)+(\vec j\times \vec r)\cdot \nabla \tilde{\vec b} -\vec j\times \tilde{\vec b} =\tilde{\Theta} \vec r + P_m\nabla^2 \tilde{\vec u}, \\
\label{103}
\partial_t \tilde{\vec b} 
- (\vec j\times \vec r)\cdot \nabla \tilde{\vec u} + \vec j\times\tilde{\vec u} = \nabla^2 \tilde{\vec b}, \\
\label{104}
\hat{R}\,\vec r \cdot \tilde{\vec u} + \nabla^2 \tilde{\Theta} -(P/P_m) \partial_t\tilde{\Theta} = 0,\\
\label{101}
\nabla \cdot \tilde{\vec u} = 0, \quad \nabla \cdot \tilde{\vec b} = 0,
\end{gather}
\end{subequations}
respectively, that govern the evolution of infinitesimal velocity
perturbations $\tilde{\vec u}$, temperature perturbations
$\tilde{\Theta}$, and magnetic field perturbations $\tilde{\vec b}$
away from the static state.
The equations have been non-dimensionalised using the radius $r_0$ as
a unit of length,  $r_0^2 / \eta$ as a unit of time, $\eta^2 / g
\alpha r_0^4$ as a unit of temperature and $\sqrt{\mu
\varrho}\eta/r_0$ as a unit of magnetic flux density.
The dimensionless magnetic field takes the form $\vec j\times \vec r +
\tilde{\vec b}$, where $\vec j=j\vec k$ is the vector of the density
of the imposed electric current. The problem is then characterised by
five dimensionless parameters, namely the Rayleigh number, the
Coriolis parameter, the Prandtl number, the magnetic Prandtl number
$P_m$ and the non-dimensional current density given by
\begin{gather}
\label{110}
\hat{R} = \frac{\alpha g \beta r_0^6}{\eta \kappa} , \enspace \tau =
\frac{2 \Omega r_0^2}{\eta} , \enspace P =
\frac{\nu}{\kappa}, \enspace P_m = \frac{\nu}{\eta},\enspace j, 
\end{gather}
respectively. {In fact, in the results obtained below the two Prandtl
numbers enter only as their ratio $S=P/P_m = \eta/\kappa$.}
To signify that in our definition of the Rayleigh number the magnetic
diffusivity replaces the kinematic viscosity we have attached a hat to
$\hat{R}$. 

\section{Perturbation analysis}

\looseness=-1
Without loss of generality we assume that the velocity, the
magnetic field and the temperature perturbations have an exponential
dependence on time $t$ and on the azimuthal angle $\phi$. Further,
since both the velocity field and the magnetic field are solenoidal we
use the poloidal-toroidal decomposition 
\begin{subequations}
\label{120}
\begin{gather}
\label{121}
\tilde{\vec u} = \vec u \exp\big(i(\omega\tau t +m\phi)\big) = \big(\nabla \times ( \nabla v \times \vec r) + \nabla w \times
\vec r\big)\exp\big(i(\omega\tau t +m\phi)\big), \\ 
\label{122}
\tilde{\vec b} = \vec b \exp\big(i(\omega\tau t +m\phi)\big) = \big(\nabla \times ( \nabla h \times \vec r) + \nabla g \times
\vec r\big)\exp\big(i(\omega\tau t +m\phi)\big), \\
\label{123}
\tilde{\Theta} = \Theta \exp\big(i(\omega\tau t +m\phi)\big).
\end{gather}
\end{subequations}
Equation \eqref{103} can now be written in the form
\begin{gather}
\vec b = \frac{m\gamma}{\omega} \vec u - \frac{i}{\omega\tau}\nabla^2 \vec b,
\end{gather}
where the parameter $\gamma$ is defined as $\gamma=j/\tau$,
and in the $\nabla$-operator the $\phi$-derivative is replaced by its
eigenfactor $im$. This allows us to transform equation
\eqref{102} in the form, 
\begin{align}
i\omega\left(1-\frac{m^2}{\omega^2}\gamma^2\right)\vec{u} &+ 
\left(1-\frac{m}{\omega}\gamma^2\right)\vec k \times \vec u  - \nabla \check{\pi}  =
\frac{1}{\tau}\Theta \vec r + \frac{P_m}{\tau}\nabla^2\vec u
+\frac{m^2\gamma^2}{\omega^2\tau}\nabla^2\vec u  \nonumber
\\
&+\frac{2m\gamma^2}{i\omega^2\tau}\vec k\times \nabla^2\vec u +
\frac{m\gamma}{\omega \tau}\nabla^2\vec b_b + \frac{2\gamma}{i\omega
  \tau}\vec k\times \nabla^2\vec b_b +  \frac{P_m}{\tau}\nabla^2 \vec u_b,  
\label{130}
\end{align}
where 
$\check{\pi}$ is the effective pressure.
In equation \eqref{130} the magnetic field $\vec b$ appears only in
the form of the boundary layer correction $\vec b_b$, required
since the basic dissipationless solution does not satisfy all boundary
conditions \citep{ZhangBusse1995}. For the same reason the Ekman layer
correction $\vec u_b$ must be introduced \citep{Zhang1994}.

Following the procedure of earlier papers
\citep{Zhang1994,BusseSimitev2004} we use a perturbation approach
and solve equation \eqref{102} in the limit of large $\tau$,  using
the ansatz
\begin{equation}
\label{190}
\vec u =\vec u_0 + \tau^{-1}\vec u_1 + ... , \quad  \omega = \omega_0
+ \tau^{-1}\omega_1 + ..., \quad
\vec b =\vec b_0 + \tau^{-1}\vec b_1 + ... , 
\end{equation}
The heat equation is solved unperturbed.

\subsection{Zeroth-order approximation}

In the following we shall assume the limit of large $\tau$ such that
in zeroth order of approximation the right hand side of equation \eqref{130}
can be neglected. The left hand side together with the condition
$\nabla\cdot \vec u=0$ is of the same form as the equation for
inertial modes \citep{Zhang1994,BusseSimitev2004}. In the nonmagnetic
case the inertial modes corresponding to the sectorial spherical
harmonics yield the lowest critical Rayleigh numbers for the onset of
convection \cite{BusseSimitev2004}. We shall assume that this property
continues to hold as long as the parameter $\gamma$ is sufficiently
small so that the nonmagnetic limit is approached in the left-hand
side of equation \eqref{130}. The sectorial inertial modes are given by 
\begin{subequations}
\begin{gather}
\label{150}
\hspace*{-0.5cm}v_0 = P_m^m( \cos \theta ) f (r), \quad 
w_0 = P_{m+1}^m ( \cos \theta ) \psi (r),
\end{gather}
with
\begin{gather}
\label{161}
f(r) = r^m - r^{m+2}, \qquad 
\psi(r) =r^{m+1} \frac{2im(m+2)}{(2m+1)(\lambda_0 (m^2+3m+2)-m)}, 
\end{gather}
where $\lambda_0$
\begin{gather}
\label{162}
\lambda_0 = \frac{1}{m+2} \left(1 \pm\sqrt{\frac{m^2+4m+3}{2m+3}} \right),
\end{gather}
\end{subequations}
is the frequency of the inertial modes. The sectorial magneto-inertial
modes are then described by the same velocity field \eqref{150} and by
a magnetic field $\vec b_0 = m\gamma \vec u_0 /\omega_0$. In the above
expressions the subscript $0$ refers to the dissipationless solution of equations
\eqref{100}. The frequency $\omega_0$ of the magneto-inertial waves is
determined  by 
\begin{equation}
\label{165}
\lambda_0 = \frac{\omega_0^2-m^2\gamma^2}{\omega_0-m\gamma^2},
\end{equation} 
which yields
\begin{gather}
\label{166}
\omega_0 = \frac{\lambda_0}{2} \pm \sqrt{\frac{\lambda_0^2}{4}+ m\gamma^2(m-\lambda_0)}.
\end{gather}
With account of \eqref{162}, this dispersion relation allows for a total
of four different frequencies $\omega_0$. For small values of
$\gamma^2$ these are given by
\begin{subequations}
\label{170}
\begin{gather}
\label{171}
\hspace*{-3mm}
\omega_{01,2} =
\frac{1}{m+2}\left(1\pm\sqrt{\frac{m^2+4m+3}{2m+3}}\right) 
  +m^2\gamma^2 (m+2)\left(1\pm\sqrt{\frac{m^2+4m+3}{2m+3}}\right)^{-1}-m\gamma^2,\\
\label{172}
\omega_{03,4} = -m^2\gamma^2 (m+2)\left(1\pm\sqrt{\frac{m^2+4m+3}{2m+3}}\right)^{-1} +m\gamma^2.
\end{gather}
\end{subequations}
The upper sign in expression \eqref{171} refers to retrogradely propagating
modified inertial waves, while the lower sign corresponds to the
progradely traveling variety. The effect of the magnetic field tends
to increase the absolute value of the frequency in both
cases. Expression \eqref{172} describes the dispersion of the slow magnetic
waves. The upper sign refers to the progradely traveling modified
Alfven waves and the lower sign corresponds to retrogradely
propagating modified Alfven waves. 

\subsection{First-order approximation}

\looseness=-1
The magneto-inertial waves described by expressions \eqref{150} satisfy the
condition that the normal component of the velocity field vanishes at
the boundary. This property implies that the normal component of the
magnetic field vanishes there as well. Additional boundary conditions
must be specified when the full dissipative problem described by \eqref{130}
is considered. We shall assume a stress-free boundary with either a
fixed temperature (case A) or a thermally insulating boundary (case B), 
\begin{equation}
\label{180}
\vec r \cdot \vec u =  \vec r \cdot \nabla (\vec r \times \vec u) /r^2
= 0 \enspace \mbox{ and } \enspace
\left\{\begin{array}{ll}
\Theta = 0  & \text{(case A)} \\
\partial_r\Theta = 0  & \text{(case B)}
\end{array}\right\} \enspace
\mbox{ at } \enspace r=1.
\end{equation}
Additionally we shall assume an electrically insulating exterior of
the sphere which requires 
\begin{equation}
g=0 \enspace \mbox{ at } \enspace r=1
\end{equation}
and the matching of the poloidal magnetic field to a potential field
outside the sphere.

After the ansatz \eqref{190} has been inserted into equation
\eqref{130} such that terms with $\vec u_1$ appear on the left hand
side, while those with $\vec u_0$ and $\omega_0$ appear on the right
hand side, we obtain the solvability condition for the equation for
$\vec u_1$ by multiplying it with  $\vec u_0^*$ and averaging it over
the fluid sphere,  
\begin{gather}
\label{200}
\hspace*{-60mm}
i\omega_1 \left\langle|\vec u_0|^2\right\rangle
 \Bigg(1+\left(\frac{m^2}{\omega_0^2}-\frac{m(\omega_0^2-m^2\gamma^2)}{\omega_0^2(\omega_0-m\gamma^2)}\right)\gamma^2\Bigg)
\\ 
\hspace*{10mm}
= \langle\Theta\vec r \cdot \vec u_0^*\rangle 
+ \left(\langle \vec u_0^* \cdot \nabla^2 \vec u_0 \rangle
  \frac{m\gamma}{\omega_0}
+ \langle \vec u_0^* \cdot \nabla^2 \vec b_{0b} \rangle
\right)
  \left(\frac{m}{\omega}-\frac{\omega_0^2-m^2\gamma^2}{\omega_0^2-m\omega_0\gamma^2}\right)
  \frac{\gamma}{\tau}, \nonumber
\end{gather}
where the brackets $\langle...\rangle$ indicate the average over the
fluid sphere and the $*$ indicates the complex conjugate. We have
neglected all terms connected with viscous dissipation, i.e. we have
assumed the vanishing of $P_m$, since we wish to focus on the effect
of ohmic dissipation. The effects of viscous dissipation have been
dealt with in the earlier paper \citep{BusseSimitev2004}.  Since
$\langle \vec u_0^* \cdot \nabla^2 \vec u_0 \rangle$ vanishes, as
demonstrated in \citep{ZhangEtal2001}, we must consider only the
influence of the boundary layer magnetic field $\vec b_{0b}$. It is
determined by the equation  
\begin{equation}
\label{210}
i\omega_0\tau\vec b_{0b}=\nabla^2\vec b_{0b}.
\end{equation} 
Since the solutions of this equation are characterized by gradients of
the order $\sqrt{\tau}$, the boundary layer correction needed for the
poloidal component is of the order $\sqrt{\tau}$ smaller than the
correction needed for the toroidal component. For large $\tau$ we need
to take into account only the contribution $g_{0b}$ given by 
\begin{align}
\label{220}
g_{0b} &= -g_0(r=1)\exp\left(-(1+is)(1-r)\sqrt{|\omega_0|\tau/2}\right) \nonumber \\
&=-\frac{m\gamma}{\omega_0}w_0(r=1)\exp\left(-(1+is)(1-r)\sqrt{|\omega_0|\tau/2}\right),
\end{align} 
where $s$ denotes the sign of $\omega_0$. The solvability condition thus becomes reduced to
\begin{align}
\label{230}
i\omega_1\left\langle|\vec u_0|^2\right\rangle 
   &\Bigg(1+\left(\frac{m\gamma^2(m-\omega_0)}{\omega_0(\omega_0-m\gamma^2)}\right)\Bigg) \\
=\, &\frac{1}{\tau}\langle\Theta\vec r \cdot \vec
u_0^*\rangle
-\frac{3}{2}
\frac{m\gamma^2(m-\omega_0)(s+i)}{(\omega_0-m\gamma^2)\sqrt{2|\omega_0|\tau}}
\int_{-1}^{1}|P_m^{m+1}|^2 \mathrm{d}( \cos\theta)\ \nonumber\\
&\times (m+1)(m+2)
\left|\frac{2m(m+2)}{(2m+1)\left(\frac{\omega_0^2-m^2\gamma^2}{\omega_0-m\gamma^2}(m+1)(m+2)-m\right)}\right|^2. \nonumber 
\end{align}

\begin{figure}[t]
\includegraphics[width=\textwidth,clip=true]{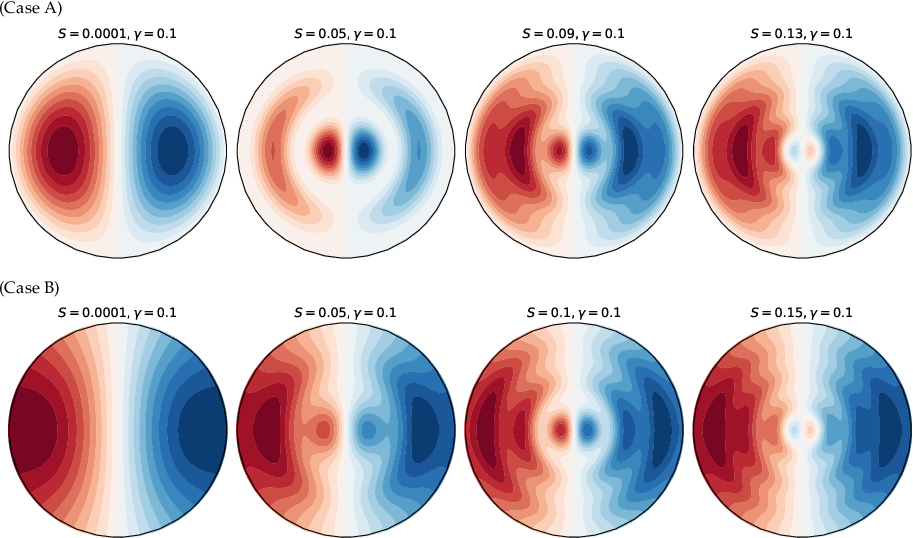}  
  \caption{Contour plots of the (normalized) temperature perturbation
    $\Theta(r)$ of the preferred mode given by equations \eqref{240}
    and \eqref{320} in case A     (top row) and case B (bottom row) with values of $S$  and $\gamma$
    as specified in the panels and $\tau=10^4$, $m=1$ and frequency $\omega_{01}$. Expressions
    \eqref{240} and \eqref{250} for the limit $\tau S \ll 1$ appear
    identical to the plots in the first column.}  
\label{contours}
\end{figure}

\begin{figure}[t]
\includegraphics[width=\textwidth,clip=true]{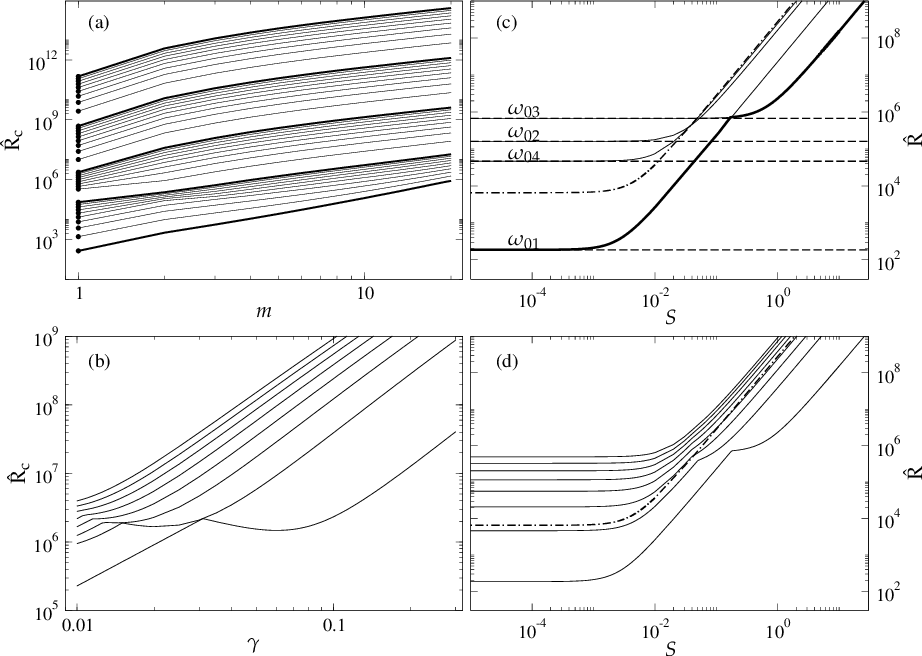}  
  \caption{Case A. 
    (a) The critical Rayleigh number $\Rc$ as a function of the wave
    number $m$ for $\gamma=0.1$ and $\tau=10^2\dots 10^6$
    increasing from bottom with log-scale decades given by the five thick
    lines. (b) The critical Rayleigh number $\Rc$ as a function of
    $\gamma$ for $S=1$ and $m=1\dots8$ increasing from
    bottom. (c) Competition of modes with increasing $S$ for 
    $\gamma=0.1$ and $m=1$. Explicit expressions \eqref{270} in the limit
    $\tau S \ll 1$ are shown by broken lines. (d) The critical
    Rayleigh number $\Rc$ as  a function of $S$ for 
    $\gamma=0.1$ and $m=1\dots8$ increasing from bottom. The
    axisymmetric mode $m=0$ is given for comparison in panels (c,d) by
    a dot-dashed line. In panels (b) to (d) $\tau=10^4$.} 
\label{f1}
\end{figure} 

\subsubsection{Explicit expressions in the limit $\tau S \ll 1$}

The equation \eqref{104} for $\Theta$ can most easily be solved in the
limit of vanishing $\omega_0 \tau S$.  In this limit we obtain for $\Theta$,
\begin{equation}
\label{240}
\Theta = P_m^m( \cos \theta ) \exp ( im \varphi + i \omega \tau t ) q(r),
\end{equation}
with
\begin{gather}
\label{250}
q(r) =\hat{R}\left(\frac{m(m+1)r^{m+4}}{(m+5)(m+4)-(m+1)m} -
  \frac{m(m+1)r^{m+2}}{(m+3)(m+2)-(m+1)m} - c r^m\right),
\end{gather}
where the coefficient $c$ is given by 
\begin{equation}
\label{260}
c = 
\begin{cases}
 \displaystyle\frac{m(m+1)}{(m+5)(m+4)-(m+1)m} -
 \displaystyle\frac{m(m+1)}{(m+3)(m+2)-(m+1)m},   & \text{case A,} \\[10pt]
 \displaystyle\frac{(m+4)(m+1)}{(m+5)(m+4)-(m+1)m} -
 \displaystyle\frac{(m+2)(m+1)}{(m+3)(m+2)-(m+1)m},  & \text{case   B.} \end{cases}
\end{equation}
Since $\Theta$ and the left hand side of equation \eqref{230} is imaginary,
the real parts of the two terms on the right hand side must
balance. We thus obtain for $\hat{R}$ the result 
\begin{gather}
\label{270}
\hspace*{-40mm}
\hat{R} = s\sqrt{\frac{\tau}{2|\omega_0|}}\frac{\gamma^2(m-\omega_0)}{(\omega_0-m\gamma^2)}
 \left|\frac{m(m+2)}{\frac{\omega_0^2-m^2\gamma^2}{\omega_0-m\gamma^2}(m+1)(m+2)-m}\right|^2 \\
\hspace*{50mm}
\times (2m+9)(2m+7)(2m+5)^2(2m+3)\frac{m+2}{m+1}\,\frac{1}{b}, \nonumber
\end{gather}
where the coefficient $b$ assumes the values 
\begin{equation}
\label{280}
b = 
\begin{cases}
 m(10m+27)   & \text{case A,} \\[2pt]
 14m^2+59m+63  & \text{case B.}
\end{cases} 
\end{equation}

Obviously the lowest value of $\R$ is usually reached for $m=1$, but
the fact that there are four different possible values of the
frequency $\omega_0$ complicates the determination of the critical
value $\Rc$. 
Expression \eqref{270} is also of interest, however, in the case of spherical
fluid shells when the ($m=1$)-mode is affected most strongly by the
presence of the inner boundary. Convection modes corresponding to
higher values of $m$ may then  become preferred at onset since their
$r$-dependence decays more rapidly with distance from the outer boundary
according to relationships \eqref{161}.    

\begin{figure}[t]
\includegraphics[width=\textwidth,clip=true]{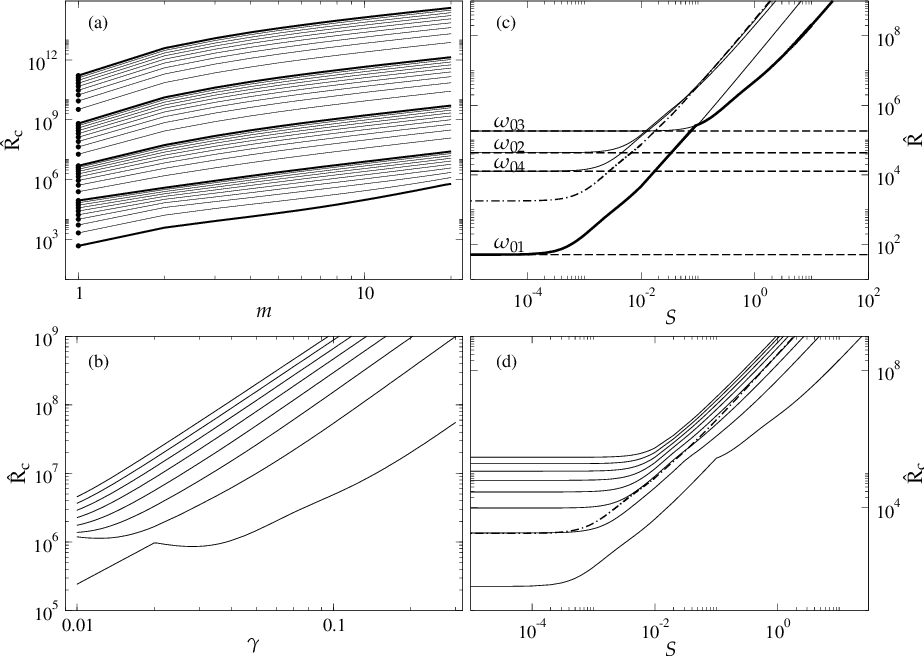}  
  \caption{Same as figure \ref{f1} but in Case B.}
\label{f2}
\end{figure}

\begin{figure}[t]
\includegraphics[width=\textwidth,clip=true]{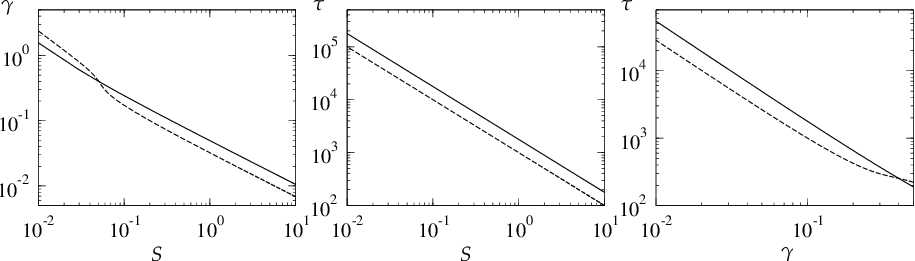}
  \caption{The boundary where the transition from modes
    characterised by $\omega_{01}$ to modes characterised by
    $\omega_{03}$ occurs in various sections of the parameter
    space. The value of the parameters are $m=1$, $S=1$,
    $\gamma=0.1$, and $\tau=5000$ where they are not varied on the
    axes. Case A is denoted by a solid lines and Case B by broken lines.}
\label{f3}
\end{figure}

\subsubsection{Solution of the heat equation in the general case}

\begin{figure}[t]
\includegraphics[width=\textwidth,clip=true]{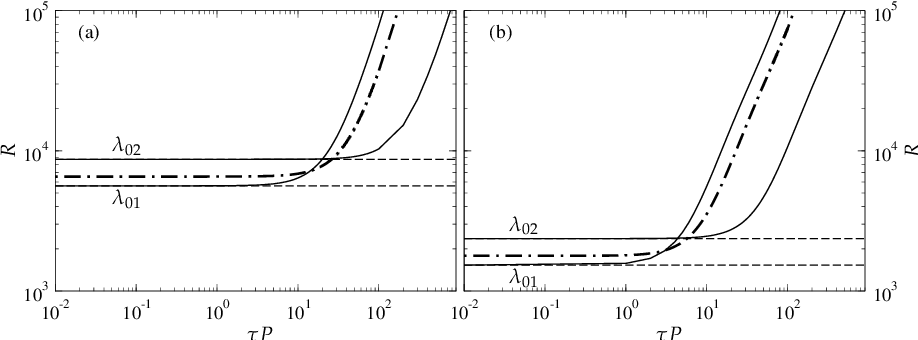}  
  \caption{
    Competition of modes with increasing $\tau P$ in the non-magnetic
    case discussed in \citep{BusseSimitev2004}. 
The Rayleigh number R as a function of $\tau P$ for $m = 0$ (thick
dash-dotted lines) and $m=1$ (thin lines). Results based on the explicit
expressions (4.6) and (3.4) from \citep{BusseSimitev2004} are shown in
solid lines and broken lines respectively in the case $m=1$. 
(a) Case A, fixed temperature boundary conditions. (b) Case B,
insulating thermal boundary conditions. }
\label{f.02}
\end{figure}

For the solution of equation \eqref{104} in the general case it is convenient
to use the Green's function method. The Green's function $G(r,a)$ is
obtained as solution of the equation 
\begin{equation}
\label{290}
 \left[\partial_r r^2 \partial_r + \big(-i \omega_0 \tau S\; r^2 - m(m+1)\big)\right] G(r,a)  = \delta(r-a),
\end{equation}
which can be solved in terms of the spherical Bessel functions $j_m(\mu
r)$ and $y_m(\mu r)$,
\begin{equation}
\label{300}
 G(r,a) = \begin{cases}
 G_1(r,a) = A_1 j_m(\mu r) \qquad & \text{for $0 \leq r < a$}, \\[2pt]
 G_2(r,a) = A j_m(\mu r)+ B y_m(\mu r)  \qquad & \text{for $a<  r \leq 1$, }
\end{cases}
\end{equation}
where
\refstepcounter{equation}
$$
\hspace*{-1.65cm}
\mu \equiv \sqrt{- i \omega_0 \tau S}, \quad
A_1= \mu\left(y_m(\mu a) - j_m(\mu a) \f{y_m(\mu)}{j_m(\mu)}
  \right), 
\eqno{(\theequation{\mathit{a},\mathit{b}})}  
$$
$$
\hspace*{-3.5cm}
  A= - \mu j_m(\mu a) \f{y_m(\mu)}{j_m(\mu)}, \quad 
  B= \mu j_m(\mu a). 
\eqno{(\theequation{\mathit{c},\mathit{d}})}  
$$
A solution of equation \eqref{104} can be obtained in the form
\begin{gather}
\label{310}
q(r) 
=- m(m+1) \R \left(\int_0^r G_2(r,a)\left(a^m-a^{m+2}\right)
a^2 \mathrm{d} a +\int_r^1 G_1(r,a)\left(a^m-a^{m+2}\right) a^2 \mathrm{d} a\right). 
\end{gather}
Evaluations of these integrals for $m=1$ yield the expressions
\begin{equation}
\label{320}
q(r) = 
\begin{cases}
\displaystyle\frac{2\R}{(\omega_0 \tau S)^2}\left( r(\mu^2 + 10) - \mu^2r^3 -
  \displaystyle\frac{10\big(\mu r\cos(\mu r) - \sin(\mu r)\big)}{r^2\big(\mu \cos\mu -
    \sin\mu \big)}\right)   & \text{case A,} \\[15pt]
\displaystyle\frac{2\R}{(\omega_0 \tau S)^2}\left( r(2\mu^2 + 10) - \mu^2r^3 -
  \displaystyle\frac{(\mu^2-10)\big(\mu r \cos(\mu r) - \sin(\mu r)\big)}{r^2\big(2\mu \cos\mu -
    (2-\mu^2)\sin\mu\big)}\right)  & \text{case B.}
\end{cases} 
\end{equation}
Lengthier expressions are obtained for $m>1$.
This first order approximation of the temperature perturbation is
illustrated in Figure \ref{contours} for the preferred modes of
inertial magnetoconvection. The preferred modes of convection at onset
are determined by minimizing the values of the critical Rayleigh number
$\hat{R}$ at given values of the other parameters. The critical
Rayleigh number $\hat{R}$ and frequency $\omega_1$ are calculated on
the basis of equation \eqref{230} using expressions \eqref{320}.
In the case $m=1$ we obtain 
\begin{align}
\label{330}
\R ={\frac {189}{20}}\,&{\frac {s \sqrt {2\tau} {{\gamma}}^{2}
      \left( \omega_0-1 \right) }{ \sqrt { \left| \omega_0
        \right|}\left( \omega_0-{{\gamma}}^{2} \right) \left(
        6\,{\lambda_0}-1 \right) ^{2}}} \\ 
&\times
\begin{cases}
\displaystyle
\left(\mu^{-4}-525\mu^{-8}-175\;\text{Re}
    \left\{\frac{\sin\mu}{\mu^6(\mu\,\cos\mu-\sin\mu)} \right\} \right) ^{-1}  & \text{case A,} \\[10pt]
\displaystyle
    \left(\mu^{-4}+231\mu^{-8}+7\;\text{Re}\left\{\frac{(\mu^5-8\mu^3+9\mu)\cos\mu-9\sin\mu}{\mu^8\big((\mu^2-2)\sin\mu+2\mu\cos\mu\big)} \right\}\right)^{-1} & \text{case B,} \\ 
\end{cases} \nonumber
\end{align}
where Re$\{\}$ indicates the real part of the term enclosed by
$\{\}$. Expressions \eqref{330} have been plotted as functions of $S$
in figures \ref{f1}(c) and \ref{f2}(c) for the cases A and B,
respectively. Four curves appear since there are four possible values
of $\omega_0$ for each $m$. For values $S$ of the order 
$10^{-2}$ or less, expressions \eqref{270} are well approached. The
retrograde mode corresponding to the positive sign in \eqref{162}
always yields the lower value of $\hat{R}$ but it looses   
its preference to the progradely traveling modified Alfven mode
corresponding to the upper sign in \eqref{172} as $S$ becomes of the
order $10^{-1}$ or larger. This transition can be understood on the 
basis of the increasing difference in phase between $\Theta$ and $u_r$
with increasing $S$. While the mode with the largest absolute
value of $\omega$ is preferred as long as $\Theta$ and $u_r$ are in
phase, the mode with the minimum absolute value of $\omega$ becomes
preferred as the phase difference increases since the latter is
detrimental to the work done by the buoyancy force.  The frequency
perturbation $\omega_1$ usually makes only a small contribution to
$\omega$ which tends to decrease the absolute value of $\omega$. This
transition shifts towards smaller values of $S$ and $\gamma$ as $\tau$
is increased as illustrated in figure \ref{f3}. The magneto-inertial
convective modes corresponding to higher values of $m=1\dots8$ exhibit
similar behaviour as figures \ref{f1}(d) and \ref{f2}(d) demonstrate for
the cases A and B, respectively. The value $m=1$ is always the
preferred value of the wavenumber, except possibly in a very narrow
range near $\gamma=0.03$ as indicated by figure \ref{f1}(a,b) in the
case A and possibly near $\gamma=0.02$ in the case B and figure
\ref{f2}(a,b). The axisymmetric mode $m=0$, given for comparison in
panels (c) and (d) in figures \ref{f1} and \ref{f2}, is never
preferred in contrast to the purely non-magnetic case where it becomes
the critical one near the transition from retrograde to prograde
inertial convection modes as seen in figure \ref{f.02}.

For very large values of $\tau$ and $S$ the Rayleigh number  $\hat{R}$
increases in proportion to $\sqrt{\tau}(\tau S)^2$ for fixed $m$. In
spite of this strong increase $\Theta$ remains of the order
$\tau^{3/2} S$ on the right hand side of equation \eqref{102}. The
perturbation approach thus continues to be valid for  $\tau
\longrightarrow \infty$ as long as $S \ll 1$ can be assumed. For any
fixed low value of $S$, however, the onset of convection in the form
of  prograde inertial modes will be replaced with increasing $\tau$ at
some point by the onset in the form of columnar magneto-convection
because the latter obeys an approximate asymptotic relationship for
$R$ of the  form $\tau^{4/3}$ (see, for example, \citet{Eltayeb1977}). This second transition depends on the value of $S$  and
will occur at higher values of $\tau$ and $R$ for lower values of
$S$. There is little chance that magneto-inertial convection occurs in
the Earth's core, for instance, since $S$ is of the order $30000$ while
the usual estimate for $\tau$ is $10^{15}$ but it might be relevant
for understanding of rapidly rotating stars with strong magnetic fields.

\section{Discussion}

A main result of the analysis of this paper is that for small values
of the magnetic Prandtl number $P_m$ and $\gamma$ an azimuthal magnetic
field exerts a stabilizing influence on the onset of convection in the
form of sectorial magneto-inertial modes. As a consequence
 magneto-convection with azimuthal wave number $m=1$ is generally
preferred at onset for both thermally-infinitely
conducting and thermally-insulating boundaries. In contrast, in the
absence of a magnetic field inertial modes with azimuthal wave number
$m=1$ are preferred but only in the case of thermally-insulating
boundaries, while in the case with infinitely conducting
thermal boundaries large  azimuthal wave numbers are preferred soon
after moderately large rotation is reached \cite{BusseSimitev2004} and
magnetic field is absent.
Axisymmetric magneto-convection is never the preferred mode at onset
while in the non-magnetic case it appears to be realized in a minute region of the parameter space only.
These results are also in contrast to  previous
magnetoconvection results obtained for larger values of $P_m$
where a destabilizing role of the azimuthal magnetic field has been found. 

The region of the parameter space investigated in the present paper
differs considerably from those analysed in previous work. Most
authors have emphasized regimes of high magnetic flux density where
the magnetic field exerts a destabilizing influence and strongly
decreases the critical Rayleigh number for onset of convection (see,
for example, \cite{Eltayeb1977,Fearn1979}). Unfortunately,
no explicitly analytical results are possible in that region of the
parameter space. Moreover the choice of parameter values has often
been motivated by applications to the problem of the geodynamo in
which case the parameter $S$ is large, perhaps as large as $10^5$, when
molecular diffusivities are used. On the other hand, small values of $S$
may be relevant for magneto-convection in stars where a high thermal
diffusivity is generated by radiation.

\authorcontributions{
Conceptualization, F.H. and R.S; formal analysis, F.H. and R.S.; data
curation, R.S.; writing--original draft preparation, F.B.;
writing--review and editing, F.H. and R.S.; visualization, R.S. funding acquisition, R.S.}  
\funding{The research of R.S. was funded by the Leverhulme Trust grant number RPG-2012-600.}
\conflictsofinterest{The authors declare no conflict of interest.}

\reftitle{References}

\end{document}